\PassOptionsToPackage{super,numbers}{natbib}
\documentclass[num-refs]{wiley-article}
\usepackage{siunitx}
\usepackage[linesnumbered,ruled,vlined]{algorithm2e}
\usepackage{amsmath}
\usepackage{multirow}
\usepackage{array}
\usepackage{booktabs}
\usepackage{graphicx}
\usepackage{caption}

\papertype{Original Article}
\paperfield{Journal Section}

\title{Advanced Chest X-Ray Analysis via Transformer-Based Image Descriptors and Cross-Model Attention Mechanism}
\author[1]{Lakshita Agarwal}
\author[2\authfn{1}]{Bindu Verma}
\contrib[\authfn{1}]{Equally contributing authors.}
\affil[1]{Department of Information Technology, Delhi Technological University, Delhi, 110042, India}
\affil[2]{Assistant Professor, Department of Information Technology, Delhi Technological University, Delhi, 110042, India}
\corraddress{Bindu Verma, Assistant Professor, Department of Information Technology, Delhi Technological University, Delhi, 110042, India}
\corremail{bindu.cvision@gmail.com}
\fundinginfo{No Funding Required}
\runningauthor{Lakshita et al.}
\begin{document}
\begin{frontmatter}
\maketitle
\begin{abstract}
The examination of chest X-ray images is a crucial component in detecting various thoracic illnesses. The study introduces a new image description generation model that integrates a Vision Transformer (ViT) encoder with cross-modal attention and a GPT-4-based transformer decoder. The ViT captures high-quality visual features from chest X-rays, which are fused with text data through cross-modal attention to improving the accuracy, context, and richness of image descriptions. The GPT-4 decoder transforms these fused features into accurate and relevant captions. The model was tested on the National Institutes of Health (NIH) and Indiana University (IU) Chest X-ray datasets. On the IU dataset, it got scores of 0.854 (B-1), 0.883 (CIDEr), 0.759 (METEOR), and 0.712 (ROUGE-L). On the NIH dataset, it had the best on all metrics: BLEU 1–4 (0.825, 0.788, 0.765, 0.752), CIDEr (0.857), METEOR (0.726), and ROUGE-L (0.705). This framework has the ability to enhance chest X-ray evaluation, assisting radiologists in more precise and efficient diagnosis.
\end{abstract}
\end{frontmatter}

\section{Introduction}
Automated image description generation is an emerging technology in medical imaging that falls under the domain of computer vision. It enables the creation of textual summaries that accurately describe visual content~\cite{beddiar2023automatic}. Chest X-ray imaging is crucial for diagnosing and treating thoracic diseases and providing important information about pulmonary and cardiovascular disorders. The integration of deep learning techniques with artificial intelligence (AI) has brought a revolution in medical image analysis in recent years~\cite{agarwal2023comparison}. These technologies enhance the precision of diagnoses and promote the efficiency of radiological workflows by expediting report generation and minimizing interpretation time~\cite{revathi2024automatic}~\cite{agarwal2024methods}. 

The proposed CrossViT-GPT4 architecture combines the cross-model attention mechanism of the Vision Transformer (ViT) with the language model of GPT-4. ViT captures spatial information at the pixel level, converts it into manageable sequences, and extracts comprehensive features through cross-model attention, which helps capture long-range dependencies. This integration of textual and visual data improves image description generation by enhancing sentence accuracy, contextual relevance, and detail, while offering scalability for diverse applications. The approach was tested on established chest X-ray datasets like NIH and IU, demonstrating its effectiveness. However, challenges such as computational complexity, limited annotated medical data, domain-specific terminology, clinical integration, and accuracy in analyzing flawed images remain. The proposed methodology combined ViT and cross-model attention with GPT-4, generating detailed annotations that improve the medical terminology as well as the spatial relationships with pixel-level information. It promises to outperform previous approaches towards improving diagnostic workflows and patient care.

The paper's contributions are as follows:
\begin{itemize}
\item The research introduces a new method, CrossViT-GPT4, that combines Vision Transformer (ViT) with cross-model attention and GPT-4 to generate detailed annotations for chest X-ray images. 
\item The ViT model with cross-model attention leverages contextual comprehension to integrate pertinent medical terminology and spatial relationships, whereas GPT-4 utilizes pixel-level information to generate sequences. 
\item The suggested model outperformed existing approaches in extensive tests conducted on benchmark datasets of the Indiana University Chest X-ray dataset and the National Institutes of Health (NIH) Chest X-ray dataset. This demonstrates the potential to enhance analysis workflows, clinical outcomes, and patient care by enabling automated production of image descriptions for better diagnosis and treatment planning.
\item Ultimately, the study offers a numerical evaluation of all methods and details of the advantages and disadvantages of the proposed model.
\end{itemize}

The paper is organized as follows: Previous research in this field that is relevant is discussed in Section~\ref{relatedwork}. Comprehensive information about the proposed method that is employed is provided in Section~\ref{methods}. The utilized datasets and the final projected descriptions and outcomes are examined in Section~\ref{results}. The work's future directions and conclusion are summarised in Section~\ref{conclusion}.

\section{Related Work}
\label{relatedwork}
The primary objective of the image description model is to identify different objects and represent their relationships through accurate, semantically correct sentences. Various methods have been adapted to medical imaging tasks~\cite{pan2022contrastive}. Sun et al.~\cite{sun2024research} proposed a feature-augmented (FA) module validated on datasets such as MS-COCO, which uses the multi-modal pre-trained CLIP model and channels attention within the encoder to enhance image description and captioning performance. Yao et al.~\cite{yao2023dual} proposed a CNN-RNN framework for anatomical structure and abnormality identification in radiological imaging reports. Li et al.~\cite{liu2021contrastive} used BERT-based models to generate radiological reports from chest X-ray images, showing the significance of contextual language understanding in medical image analysis. Shaikh et al.~\cite{shaikh2023transformer} proposed an encoder-decoder transformer model combined with a pre-trained CheXNet model, evaluated on the IU X-ray dataset, for chest X-ray report generation. The CheXReport model~\cite{zeiser2024chexreport} obtains the best-reported performance on the MIMIC-CXR dataset, given the use of Swin Transformer blocks. Retrieval-based methods that incorporate deep neural networks are also presented ~\cite{reale2024vision}. In another work, Conditional Self Attention Memory-Driven Transformer was proposed, which outperformed all existing state-of-the-art approaches with a high BLEU score value for radiological report production by taking ResNet152 v2 for feature extraction and the self-attention memory-driven transformer for text generation~\cite{shahzadi2024csamdt}. Despite these developments, current models fail to integrate textual and visual information effectively and are not able to handle medical language well, which makes the descriptions less understandable and accurate. Moreover, because these models are highly sensitive to errors, misunderstandings in medical contexts could potentially affect diagnoses.

The suggested approach, CrossViT-GPT4, enhances previous research by integrating the benefits of GPT-4, which excels in contextual language modeling, with ViT and a cross-model attention technique for spatial feature extraction. This combination offers a comprehensive solution for automatically generating image descriptions in chest X-ray analysis.

\section{Overview of the Proposed Work}
\label{methods}
\begin{figure*}[!ht]
\centering
\includegraphics[width=1\textwidth]{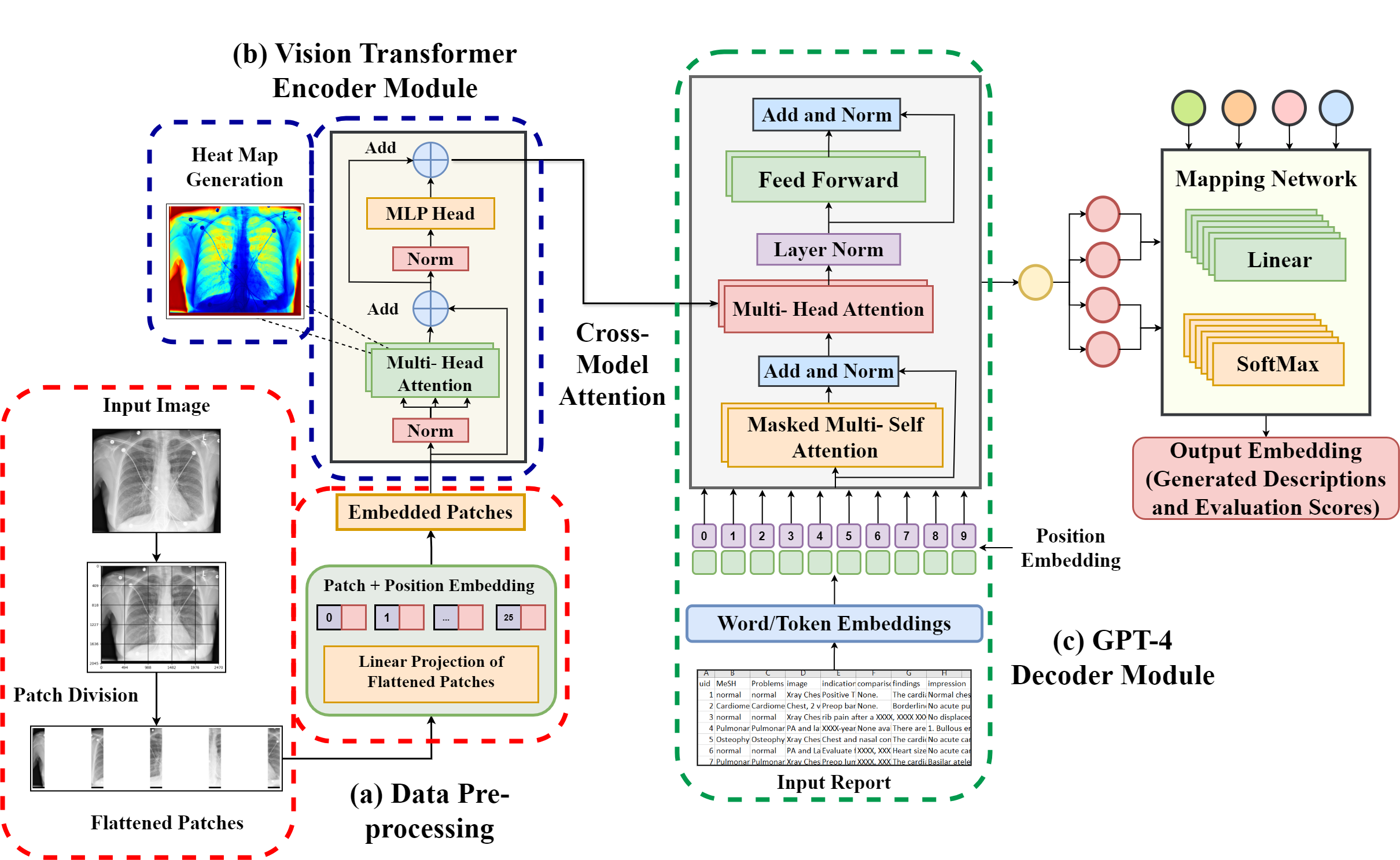}
\caption{A representation of the proposed framework's structure: CrossViT-GPT4- During the initial stage, the ViT encoder utilizes cross-model attention to extract high-level visual features from the pre-processed images. GPT-4 decoder uses tokenization to extract individual words from the provided input caption file. It then utilizes the dense layers of the network to produce image descriptions.}
\label{architecture}
\end{figure*}

The automated generation of image descriptions has received recent attention due to its intricate nature, which integrates natural language processing (NLP) with computer vision. The objective of the proposed research is to enhance the extraction of spatial and channel-wise information from images. The model is fine-tuned over multiple epochs (75) using the cross-entropy loss function and the Adam optimizer. Figure~\ref{architecture} represents the overall framework of the proposed approach. 

Each step of the proposed model is discussed below:
\subsection{Data pre-processing}
The input images ($I$) are initially partitioned into patches ($P$) of a predetermined size to prepare the input data for the encoder. The patches undergo flattening and linear projection, and each individual patch is thereafter embedded and compressed into a sequence of vectors. The data pre-processing step for the task is shown in Figure~\ref{architecture}(a). Equation~\eqref{patch} denotes compressing each patch into a lower-dimensional space:

\begin{equation}
\label{patch}
E_{\text{patch}} = \text{Linear}(\text{Flatten}(I_{\text{patch}})) + \text{PE}
\end{equation}

where, \( I_{\text{patch}} \) represents the image patches, \text{Linear} denotes the linear transformation matrix used for embedding, and \text{Flatten} is the operation that flattens each patch into a vector. The positional encoding (\text{PE}) uses sine and cosine functions to include spatial information. This lets the model understand the order of events, as shown in Equations~\eqref{pos} and~\eqref{pos1}:
\begin{equation}
\label{pos}
\text{PE}(\text{pos}, 2i) = \sin\left(\frac{\text{pos}}{10000^{(2i/\text{dim})}}\right)
\end{equation}

\begin{equation}
\label{pos1}
\text{PE}(\text{pos}, 2i+1) = \cos\left(\frac{\text{pos}}{10000^{(2i/\text{dim})}}\right)
\end{equation}

\subsection{Encoder Module of the Proposed Architecture}
The encoder part efficiently collects the spatial and visual characteristics of the image, thereby preparing the model for subsequent tasks such as image labeling. The vision transformer (ViT) encoder is used to receive the sequences of tokenized patches that have been enhanced with positional data. The architectural structure for the Vision transformer encoder is illustrated in Figure~\ref{architecture}(b). For multi-head self-attention, the attention is computed in parallel across $h$ heads and concatenated, as denoted in Equations~\eqref{attention},~\eqref{multi} and~\eqref{head}:
\begin{equation}
\label{attention}
\text{Attention}(Q, K, V) = \text{Softmax}\left(\frac{QK^T}{\sqrt{d_k}}\right) V
\end{equation}

\begin{equation}
\label{multi}
\text{MultiHead}(Q, K, V) = \text{Concat}\left(\text{Head}_1, \ldots, \text{Head}_h\right) W^O
\end{equation}

\begin{equation}
\label{head}
\text{Head}_i = \text{Attention}(Q W_i^Q, K W_i^K, V W_i^V)
\end{equation}

where, \( Q \), \( K \), and \( V \) represent the queries, keys, and values matrices, respectively. \( d_k \) is the dimension of the keys, \( W_i^Q \), \( W_i^K \), and \( W_i^V \) are projection matrices for each head and \( W^O \) is the output projection matrix. The multi-head attention module further generates heat maps to offer insights into the attention mechanism of the model, as illustrated in Figure~\ref{architecture}(b). These heat maps aid in identifying the specific areas of the input image that impact key description elements. Additionally, the cross-model attention mechanism is used to transmit these extracted features to the GPT-4 decoder module for final mapping.

\subsection{Cross-model attention mechanism}
Through cross-modal attention, which dynamically focuses on pertinent features from each modality, the proposed model is able to match visual details with textual descriptions. In image description generation, the model creates sentences by focusing on various aspects of an image and lining them up with the context that the text provides. Cross-model attention mechanism can be represented by the following Figure~\ref{crossmodel}.
\begin{figure}[!ht]
    \centering
    \includegraphics[width=0.75\textwidth]{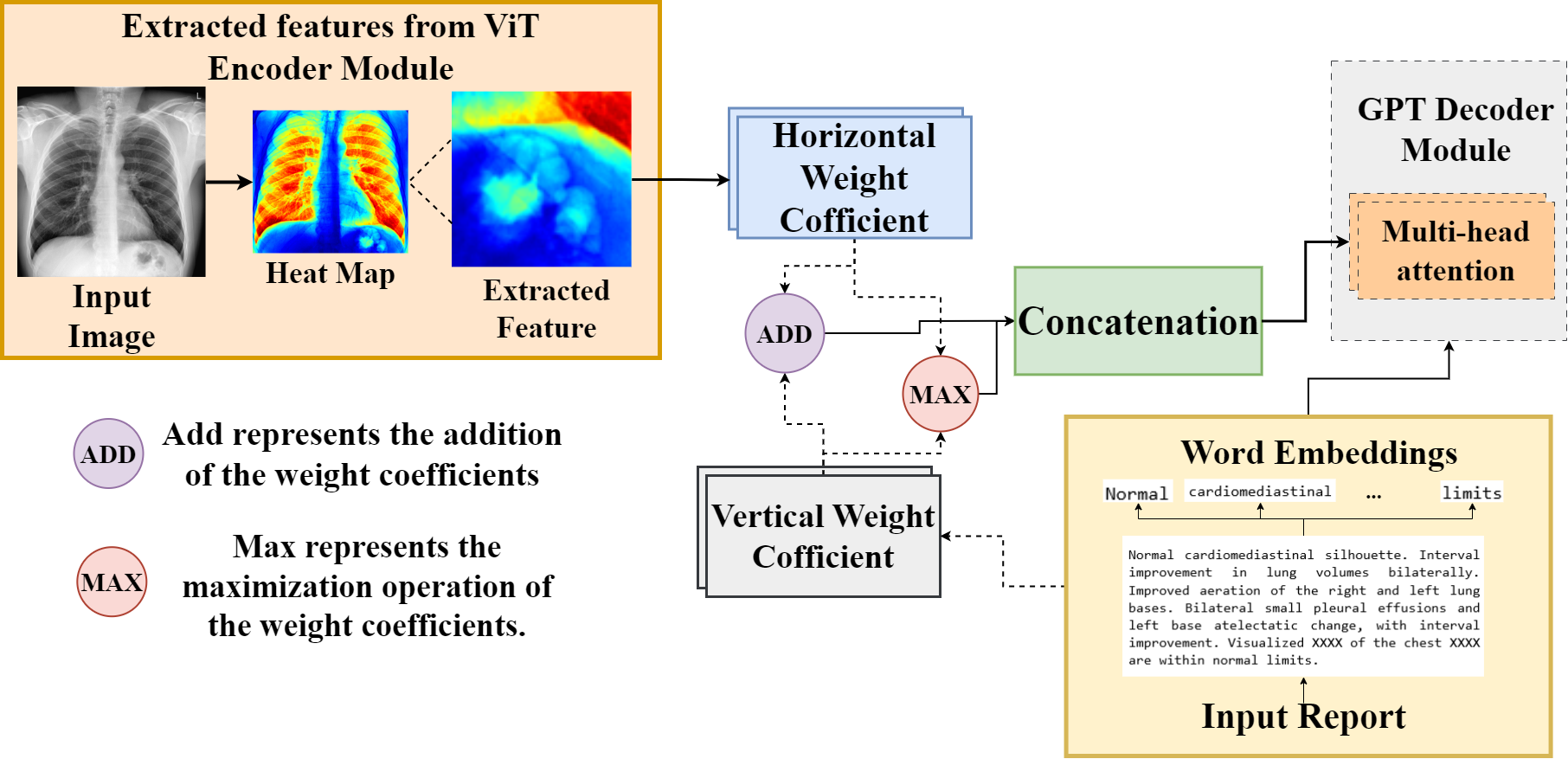}
    \caption{Cross-Model Attention Mechanism.}
    \label{crossmodel}
\end{figure}
The alignment enhances the model's capacity to generate precise and contextually appropriate descriptions, eventually leading to a richer understanding and more coherent outputs. The cross-modal attention mechanism is defined as Equation~\eqref{cross}:
\begin{equation}
\label{cross}
\text{CrossAttention}(Q_{\text{t}}, K_{\text{i}}, V_{\text{i}}) = \text{Softmax}\left(\frac{Q_{\text{t}} K_{\text{i}}^T}{\sqrt{d_{\text{t}}}}\right) \\ V_{\text{i}}
\end{equation}

where, \( Q_{\text{t}} \) represents text queries, \( K_{\text{i}} \) and \( V_{\text{i}} \) are the keys and values from the image features and \( d_{\text{t}} \) is the dimension of the text queries.

\subsection{Decoder Module of the Proposed Architecture}
The primary architecture of the decoder is based on the GPT-4 transformer model, which is specifically designed for context-based tasks. The structure consists of layers of feed-forward and multiple self-attention neural networks. The diagram in Figure~\ref{architecture}(c) illustrates the structural framework of GPT-4. Using self-attention mechanisms, every layer progressively enhances input embedding, enabling the model to discern distant relationships and contextual information. The self-attention within the GPT-4 decoder is given by Equation~\eqref{selfgpt4}:
\begin{equation}
\label{selfgpt4}
\text{SelfAttention}(Q_{\text{t}}, K_{\text{t}}, V_{\text{t}}) = \text{Softmax}\left(\frac{Q_{\text{t}} K_{\text{t}}^T}{\sqrt{d_{\text{t}}}}\right) V_{\text{t}}
\end{equation}

where, \( Q_{\text{t}} \), \( K_{\text{t}} \), and \( V_{\text{t}} \) represent the text queries, keys, and values respectively and \( d_{\text{t}} \) is the dimension of the text queries. To create the next word in the sequence, the output layer predicts the probability distribution across the vocabulary using word/token embedding, representing words as vectors. Due to this methodology, GPT-4 can generate cohesive and contextually suitable information by extensively working on more substantial sentences. The final textual output is obtained by passing the combined attention outputs through a feed-forward (\text{FF}) network and then generating Logits for each token in the vocabulary as represented in Equation~\eqref{logits}:
\begin{equation}
\label{logits}
\text{Logits} = \text{FF}\left( \text{CrossAttention} + \text{SelfAttention}\right)
\end{equation}

Linear and Softmax layers in network mapping are essential for producing the end result and predicting evaluation scores when generating image descriptions. The model can represent intricate relationships with them appropriately. Equation~\eqref{linear} represents the expression denoted as $\text{Linear}(W, X)$.
\begin{equation}
\label{linear}
\text{Linear}(W, X) = WX + b
\end{equation}

The weight matrix is symbolized by $W$, the input vector is indicated by $X$, and the bias vector is designated by $b$. Next, the updated characteristics are normalized into a probability distribution throughout the lexicon of potential words or tokens using the Softmax layer, as represented by Equation~\ref{softmax}.
\begin{equation}
\label{softmax}
\text{Softmax}(\mathbf{z})_i = \frac{e^{z_i}}{\sum_{j} e^{z_j}},
\end{equation}

where, the input vector is denoted as $z$. The accuracy of these predictions is evaluated by comparing them to the descriptions in the reference file. The evaluation metrics used are ROUGE-L, CIDEr, METEOR, and BLEU 1-4, which assess the degree of similarities between the predicted and actual descriptions. These metrics enable the evaluation of the model's ability to effectively integrate the context and semantics of the images into the generated descriptions.

\subsection{Algorithm of the Proposed Work:}
The following describes Algorithm~\ref{algo} for the suggested model along with its score evaluation:
\begin{algorithm}[!ht]
\caption{Algorithm for the proposed model, CrossViT-GPT4}
\label{algo}
\SetAlgoLined
\KwIn{Images $I$}
\KwOut{Generated Descriptions $\text{Output}_{\text{text}}$, Evaluation Scores}

\textbf{Data Pre-processing:}\\
Partition the image $I$ into patches $P$, flatten and linearly project each patch\;
Apply positional encoding to patches\;

\textbf{ViT Encoding:}\\
Input the sequence of embedded patches $E_{\text{patch}}$ into the ViT encoder\;
Compute multi-head attention: $\text{MultiHead}(Q, K, V)$\;

\textbf{Cross-Modal Attention:}\\
Compute cross-modal attention: $\text{CrossAttention}(Q_{\text{t}}, K_{\text{i}}, V_{\text{i}})$\;

\textbf{GPT-4 Decoding:}\\
\For{each time step $t$ in decoding}{
    Compute self-attention within the text: $\text{SelfAttention}(Q_{\text{t}}, K_{\text{t}}, V_{\text{t}})$\;
    Generate logits for the next token: \text{Logits}\;
    Compute probability distribution and select the next token: $\text{Output}_{\text{text}} = \text{Softmax}(\text{Logits})$\;
}

\textbf{Evaluation:}\\
Evaluate generated descriptions using BLEU, CIDEr, METEOR, and ROUGE-L scores\;

\end{algorithm}

\section{Experimental Analysis}
\label{results}
The experiments were performed using Google Colab Pro+, which provides high-performance resources, including 52 GB of RAM, an NVIDIA A100 GPU with 40 GB of VRAM, and a virtual CPU equivalent to an Intel Xeon processor. The framework was implemented using Keras and TensorFlow 2.12. The ViT model has 86.7 million parameters and GPT-4 has 1.76 trillion. Input images are (batch\_size, 224, 224, 3), and output shapes vary by layer. The developed framework was evaluated on two benchmark datasets: the Indiana University Chest X-Ray dataset (IU X-Ray) and the NIH Chest X-ray dataset. The IU X-ray dataset consists of 3,955 XML radiologist reports and 7,471 PNG images while the NIH dataset consists of 112,120 frontal X-rays with annotations for 13 thoracic illnesses across 30,805 individuals.

\subsection{Ablation Study:}
The suggested model's image description components were clarified using ablation analysis. This study investigates the capabilities of the proposed architecture, CrossViT-GPT4. In Table~\ref{results_table}, we compare the performance of various models on two major chest X-ray datasets: the IU Chest X-Ray dataset and the NIH Chest X-Ray dataset. The results presented in the following table summarize the performance of each model across these metrics.
\begin{table*}[!ht]
\centering
\caption{Evaluation of Models on the IU and NIH Chest X-Ray Datasets}
\label{results_table}
\resizebox{\textwidth}{!}{
\begin{tabular}{|l|p{3cm}|c|c|c|c|c|c|c|}
\hline
\textbf{Dataset} & \textbf{Model} & \textbf{B-1} & \textbf{B-2} & \textbf{B-3} & \textbf{B-4} & \textbf{C} & \textbf{M} & \textbf{R-L} \\
\hline
\hline
\multirow{7}{*}{IU Chest X-Ray} 
& CNN-LSTM Model & 0.289 & 0.272 & 0.268 & 0.236 & 0.677 & 0.273 & 0.219 \\
& Encoder-Decoder Transformer & 0.398 & 0.373 & 0.354 & 0.343 & 0.658 & 0.286 & 0.316 \\
& ViT+ Attention & 0.283 & 0.258 & 0.236 & 0.205 & 0.574 & 0.194 & 0.188 \\
& ViT + attention + LSTM & 0.601 & 0.572 & 0.554 & 0.498 & 0.587 & 0.436 & 0.342 \\
& ViT + GPT 2.0 & 0.704 & 0.663 & 0.634 & 0.626 & 0.792 & 0.564 & 0.474 \\
& CNN + GPT 3.0 & 0.761 & 0.753 & 0.736 & 0.715 & 0.867 & 0.725 & 0.657 \\
& \textbf{CrossViT-GPT4 (Proposed)} & \textbf{0.854} & \textbf{0.817} & \textbf{0.804} & \textbf{0.785} & \textbf{0.883} & \textbf{0.759} & \textbf{0.712} \\
\hline
\multirow{7}{*}{NIH Chest X-Ray} 
& CNN-RNN Model & 0.211 & 0.178 & 0.161 & 0.148 & 0.445 & 0.198 & 0.149 \\
& CNN-LSTM Model & 0.277 & 0.252 & 0.246 & 0.227 & 0.465 & 0.242 & 0.197 \\
& Encoder-Decoder Transformer & 0.383 & 0.363 & 0.334 & 0.312 & 0.578 & 0.265 & 0.287 \\
& ViT with Attention & 0.284 & 0.268 & 0.253 & 0.248 & 0.534 & 0.187 & 0.176 \\
& ViT with LSTM & 0.438 & 0.425 & 0.409 & 0.387 & 0.589 & 0.436 & 0.386 \\
& ViT + GPT 2.0 & 0.687 & 0.658 & 0.644 & 0.612 & 0.763 & 0.582 & 0.604 \\
& CNN + GPT 3.0 & 0.782 & 0.774 & 0.752 & 0.706 & 0.835 & 0.704 & 0.664 \\
& \textbf{CrossViT-GPT4 (Proposed)} & \textbf{0.825} & \textbf{0.806} & \textbf{0.795} & \textbf{0.772} & \textbf{0.857} & \textbf{0.726} & \textbf{0.705} \\
\hline
\end{tabular}}
\end{table*}

The table presents the performance comparison of different image description generation models on IU and NIH Chest X-ray datasets with respect to BLEU scores from B-1 to B-4, CIDEr (C), METEOR (M), and ROUGE-L (R-L). The proposed CrossViT-GPT4 model is seen to outperform all other models for both the datasets with respect to all metrics. It reports a BLEU score of 0.854 on B-1, 0.817 on B-2, 0.804 on B-3, and 0.785 on B-4 along with CIDEr of 0.883, METEOR of 0.759, and ROUGE-L of 0.712 on the IU Chest X-ray dataset. Correspondingly, it has a score of 0.825 on B-1, 0.806 on B-2, 0.795 on B-3, 0.772 on B-4, CIDEr of 0.857, METEOR of 0.726, and ROUGE-L of 0.705 on the NIH Chest X-ray dataset.

\subsection{Results and Analysis:}
The suggested model, CrossViT-GPT4, provides an excellent approach for illustrating images in the IU and NIH chest X-ray datasets. It combines the ViT encoder and the GPT-4 decoder modules. The results for the chest X-ray datasets from IU and NIH are shown in able~\ref{iuresult} and~\ref{nihresult}, respectively.
\begin{table}[!ht]
\centering
\caption{Quantitative Results Obtained on IU Chest X-Ray Dataset}
\label{iuresult}
\begin{tabular}{|p{2.5cm}|p{5cm}|p{2.5cm}|p{2cm}|}
\hline
\textbf{Test Image} & \textbf{Ground Truth Report} & \textbf{Disease Prediction} & \textbf{Predicted Report} \\
\hline
\hline
\parbox[c][2cm][c]{2.5cm}{\includegraphics[width=2.5cm]{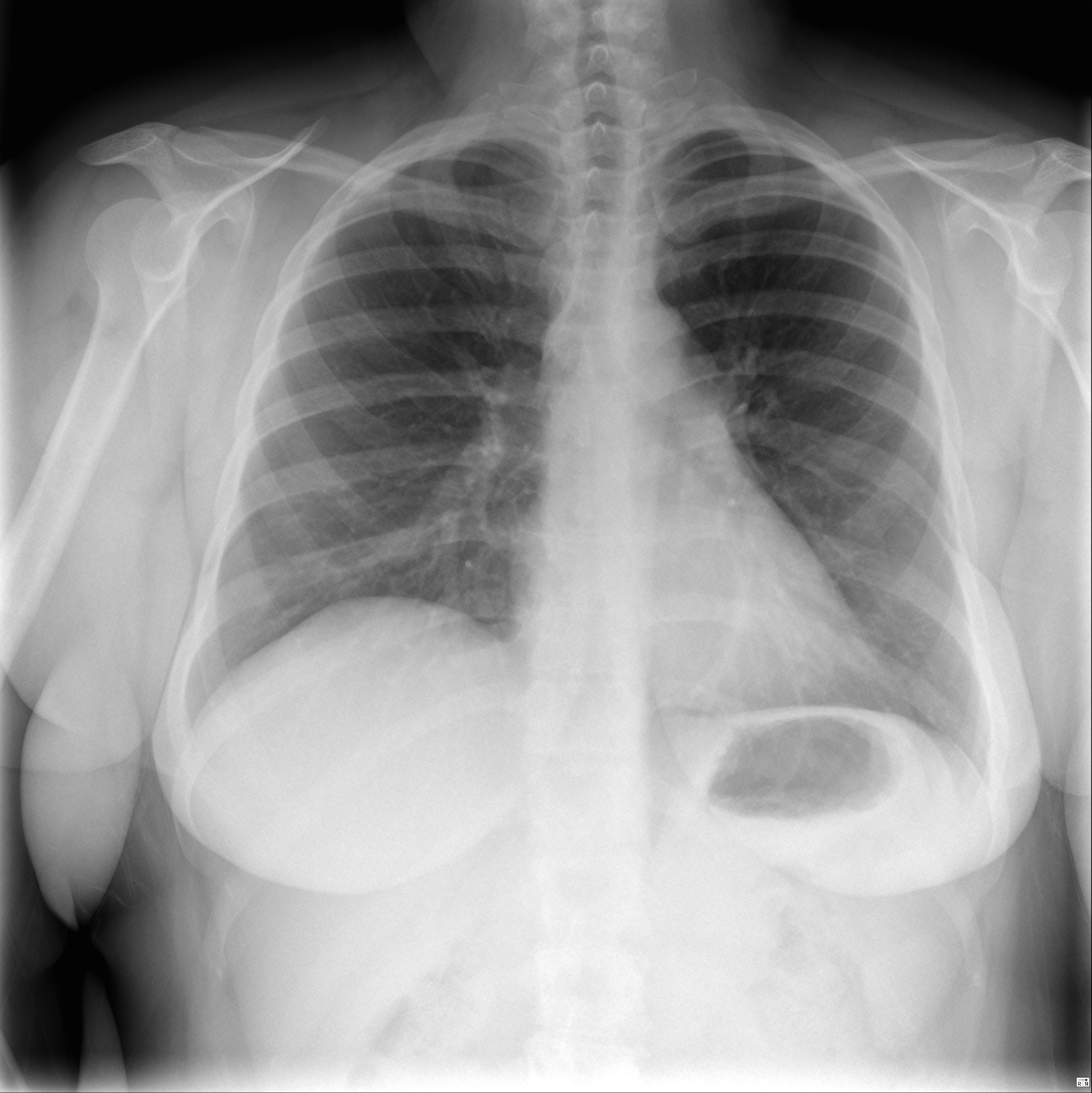}} & 
\parbox[c][3.5cm][c]{5cm}{The heart is normal in size. The mediastinal contours are within normal limits. There is mild prominence of the superior mediastinum which is somewhat lucent and reflects mediastinal and vascular structures. No focal consolidation is seen. There is no pleural effusion.} & 
\parbox[c][1.5cm][c]{2.5cm}{\includegraphics[width=2.5cm]{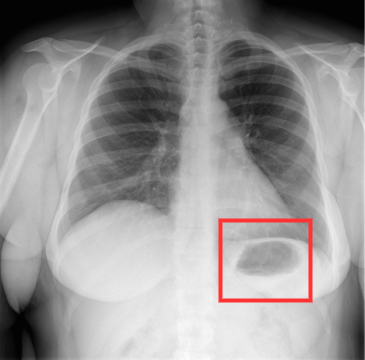}} & 
\parbox[c][2cm][c]{2cm}{Mediastinal contours within normal limits. Mild mediastinum somewhat lucent.} \\
\hline
\parbox[c][1.7cm][c]{2.5cm}{\includegraphics[width=2.5cm]{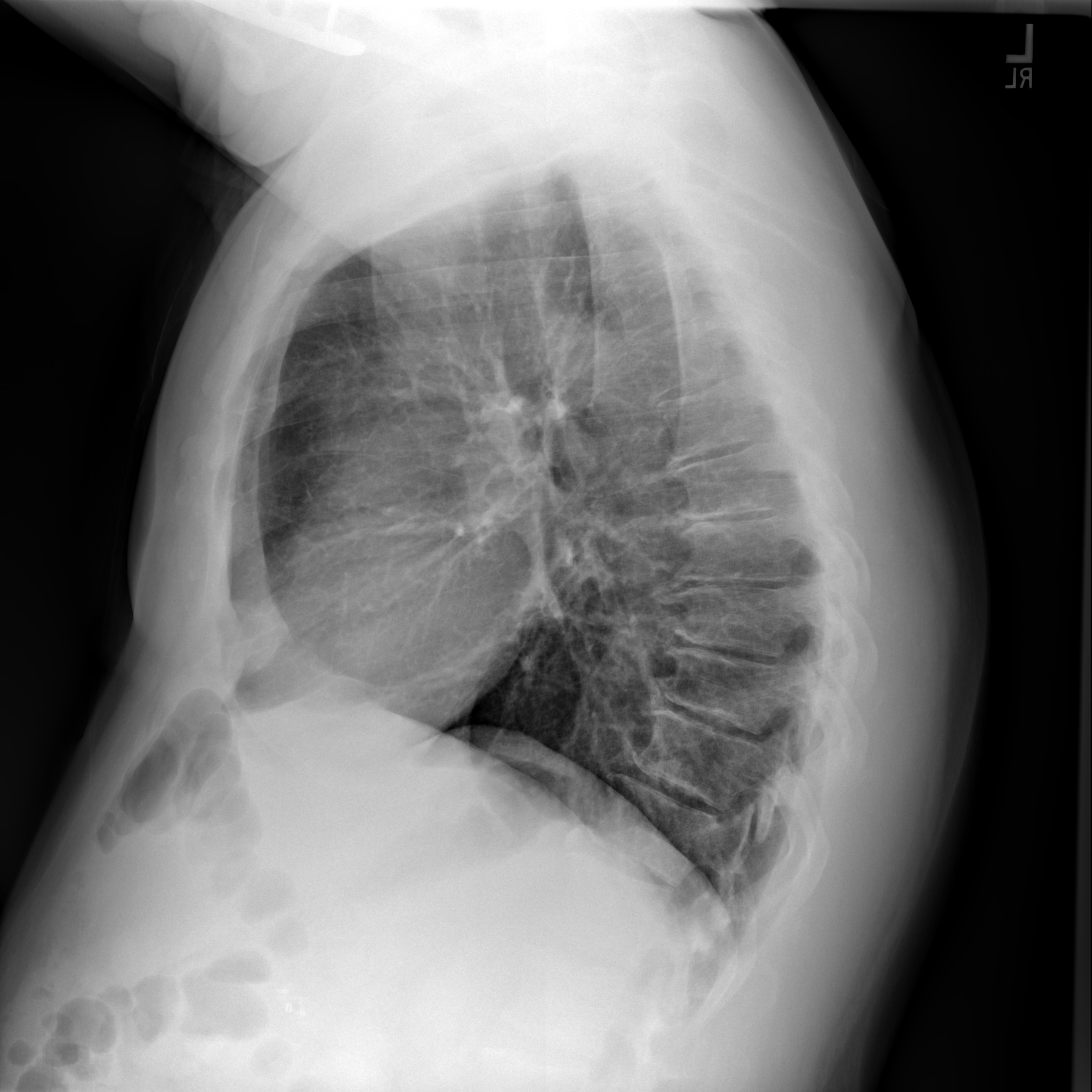}} & 
\parbox[c][3.5cm][c]{5cm}{Lateral view of the chest shows an unchanged cardiomediastinal silhouette. The cardiac silhouette remains moderately enlarged, exaggerated by epicardial fat pads. Interstitium is prominent. No focal airspace consolidation or pleural effusion. There is spine spondylosis.} & 
\parbox[c][1.5cm][c]{2.5cm}{\includegraphics[width=2.5cm]{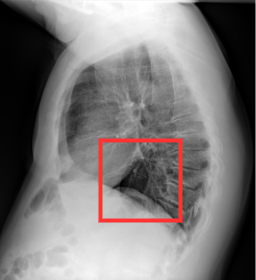}} & 
\parbox[c][1.7cm][c]{2cm}{Unchanged cardiomediastinal silhouette. Interstitium prominent. Spine spondylosis.} \\
\hline
\end{tabular}
\end{table}

\begin{table}[!ht]
\centering
\caption{Quantitative Results Obtained on NIH Chest X-Ray Dataset}
\label{nihresult}
\begin{tabular}{|p{2.5cm}|p{5cm}|p{2.5cm}|p{2cm}|}
\hline
\textbf{Test Image} & \textbf{Ground Truth Report} & \textbf{Disease Prediction} & \textbf{Predicted Report} \\
\hline
\hline
\parbox[c][2cm][c]{2.5cm}{\includegraphics[width=2.5cm]{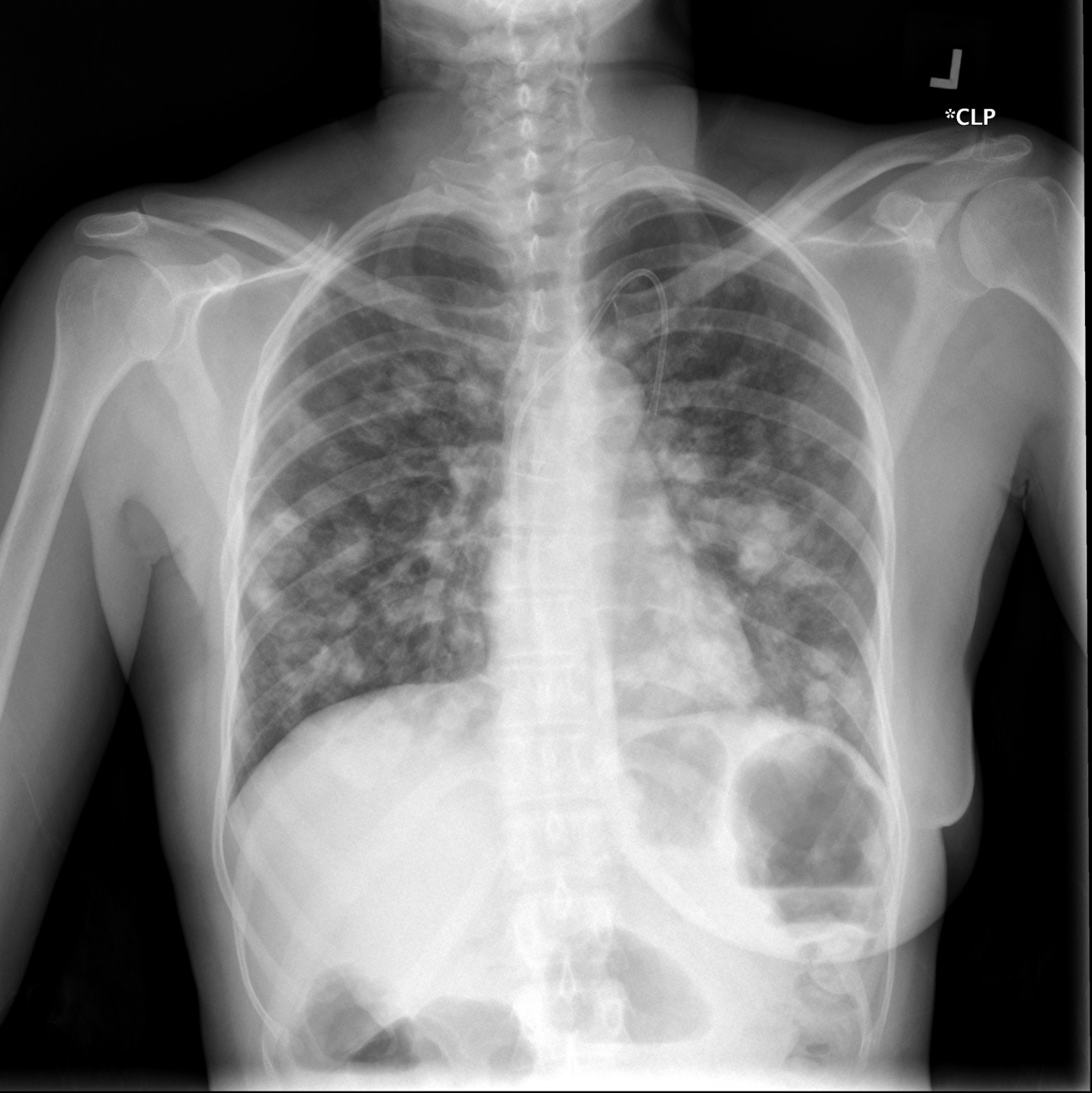}} & 
\parbox[c][3.2cm][c]{5cm}{1. The heart is normal size. 2. The mediastinum is unremarkable. 3. There is no pleural effusion, pneumothorax, or focal airspace disease. 4. There is stable irregularity of the posterior left 6th rib which represents an old fracture.} & 
\parbox[c][1.5cm][c]{2.5cm}{\includegraphics[width=2.5cm]{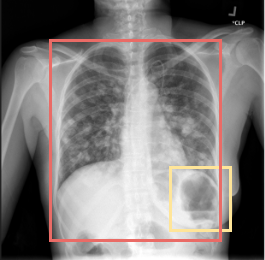}} & 
\parbox[c][2cm][c]{2cm}{No pleural effusion, pneumothorax, or focal airspace disease} \\
\hline
\parbox[c][1.7cm][c]{2.5cm}{\includegraphics[width=2.5cm]{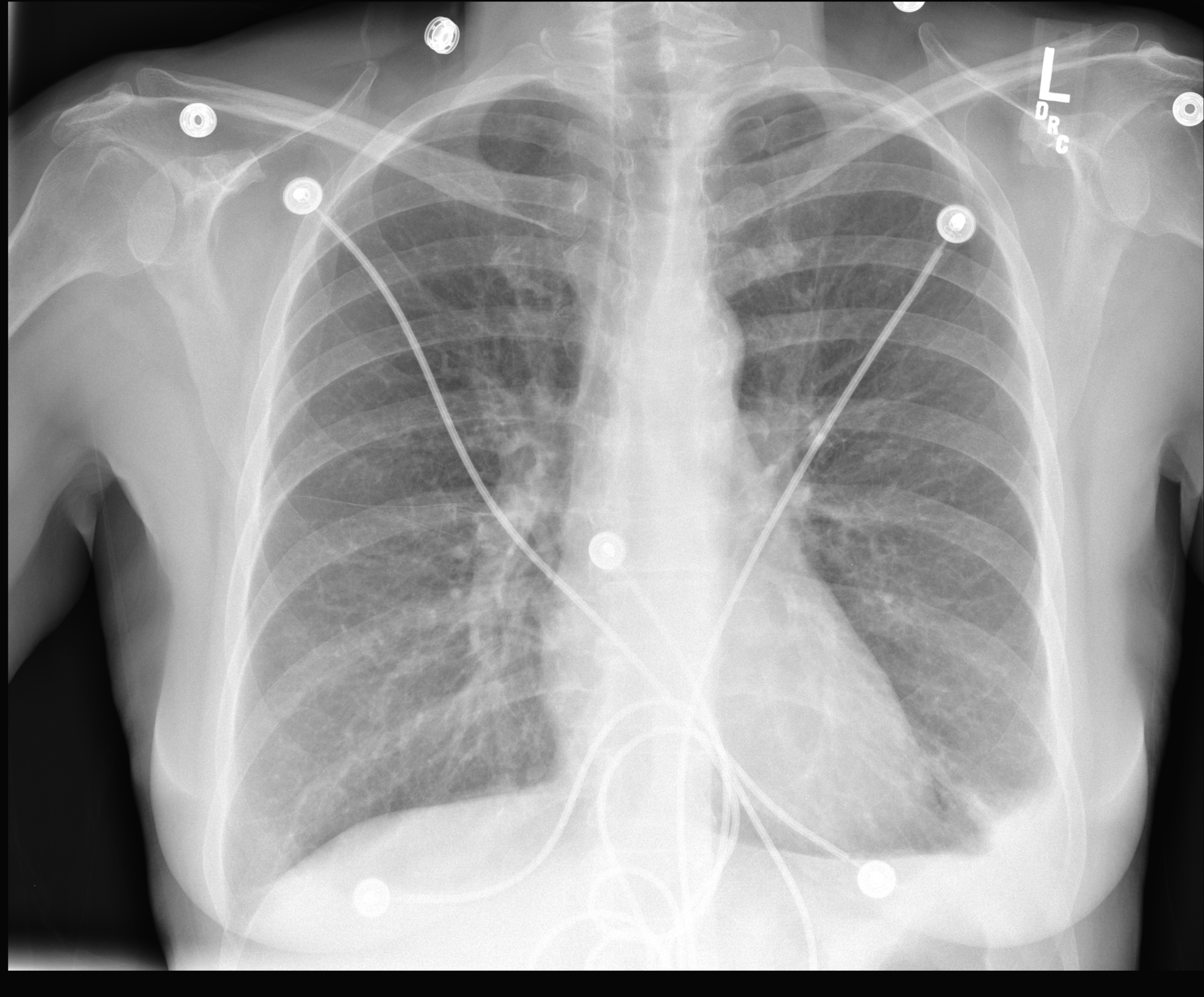}} & 
\parbox[c][3.2cm][c]{5cm}{1. The heart size and cardiomediastinal silhouette are normal. 2. There is hyperexpansion of the lungs with flattening of the hemidiaphragms. 3. There is no focal airspace opacity, pleural effusion, or pneumothorax. 4. There multilevel degenerative changes of thoracic spine.} & 
\parbox[c][1.5cm][c]{2.5cm}{\includegraphics[width=2.5cm]{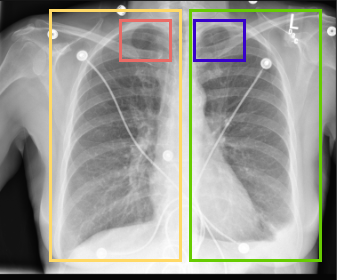}} & 
\parbox[c][1.7cm][c]{2cm}{Cardiovascular silhouette normal. No focal airspace opacity, pleural effusion, or pneumothorax.} \\
\hline
\end{tabular}
\end{table}

The tables provide a quantitative comparison between the actual data and the reports generated by the model for different test images. The results evaluate the model's capacity to forecast diseases and produce descriptions that showcase the model's precision by utilizing X-ray images for both datasets.

\subsection{Comparison With Other state-of-the-art Methods:}
Compared to the most advanced methodologies currently available, the proposed method demonstrates a substantial enhancement in disease prediction accuracy and report production for X-ray images. The method,  CrossViT-GPT4, effectively performs on both the IU and NIH chest X-ray datasets by utilizing a fusion model that incorporates the ViT encoder and GPT-4.0 decoder. 

The study presented in Table~\ref{comp_iu_nih} thoroughly analyzes various models used to generate image descriptions on the IU X-ray and NIH datasets.
\begin{table}[!ht]
\centering
\caption{Comparative Analysis on IU and NIH Chest X-Ray Datasets}
\begin{tabular}{|l|c|c|c|c|}
\hline
\textbf{Model} & \textbf{BLEU} & \textbf{METEOR} & \textbf{CIDEr} & \textbf{Rouge-L} \\
\hline
\hline
\multicolumn{5}{|c|}{\textbf{IU Chest X-Ray Dataset}} \\
\hline
R2Gen~\cite{chen2020generating} & 0.470 & 0.187 & - & 0.371 \\
R2Gen + ChexNet~\cite{wijerathna2022chest} & 0.508 & 0.222 & - & 0.365 \\
Cross-modal PROtotype driven NETwork (XPRONET)~\cite{wang2022cross} & 0.525 & 0.220 & - & 0.411 \\
Contrastive attention~\cite{liu2021contrastive} & 0.492 & 0.193 & - & 0.381 \\
Knowledge-injected U-Transformer~\cite{huang2023kiut} & 0.525 & 0.242 & - & 0.409 \\
AERMNet~\cite{zeng2024aermnet} & 0.486 & 0.219 & 0.560 & 0.398 \\
\textbf{CrossViT-GPT4 (Proposed)} & \textbf{0.854} & \textbf{0.759} & \textbf{0.883} & \textbf{0.712} \\
\hline
\hline
\multicolumn{5}{|c|}{\textbf{NIH Chest X-Ray Dataset}} \\
\hline
Semantic Attention~\cite{you2016image} & 0.467 & 0.192 & 0.560 & 0.204 \\
Co-Attention~\cite{jing2017automatic} & 0.756 & 0.597 & 0.755 & 0.675 \\
Clinical-BERT~\cite{yan2022clinical} & 0.383 & 0.144 & - & 0.275 \\
ChestBioX-Gen~\cite{ouis2024chestbiox} & 0.668 & 0.189 & 0.416 & 0.674 \\
\textbf{CrossViT-GPT4 (Proposed)} & \textbf{0.825} & \textbf{0.726} & \textbf{0.857} & \textbf{0.705} \\
\hline
\end{tabular}
\label{comp_iu_nih}
\end{table}
The table compares various models for generating image descriptions on the IU Chest X-Ray and NIH Chest X-Ray datasets, evaluating their performance using BLEU, METEOR, CIDEr, and Rouge-L metrics. On the IU Chest X-Ray dataset, R2Gen~\cite{chen2020generating} demonstrates intermediate performance, while R2Gen + ChexNet~\cite{wijerathna2022chest} improves BLEU and METEOR scores but slightly reduces Rouge-L. Both XPRONET~\cite{wang2022cross} and Knowledge-injected U-Transformer~\cite{liu2021contrastive} achieve a BLEU score of 0.525, with the latter excelling in METEOR and Rouge-L. Contrastive Attention exhibits moderate performance, whereas AERMNet~\cite{zeng2024aermnet} provides balanced results across metrics. The proposed CrossViT-GPT4 model outperforms all these approaches, achieving significantly higher scores across all metrics: BLEU (0.854), METEOR (0.759), CIDEr (0.883), and Rouge-L (0.712). On the NIH Chest X-Ray dataset, the CrossViT-GPT4 model similarly leads with the highest scores across all metrics: BLEU (0.825), METEOR (0.726), CIDEr (0.857), and Rouge-L (0.705). In comparison, Semantic Attention~\cite{you2016image}, Co-Attention~\cite{jing2017automatic}, ChestBioX-Gen~\cite{ouis2024chestbiox}, and Clinical-BERT~\cite{yan2022clinical} exhibit less balanced or lower performance. These results underscore the superior capability of the proposed model in generating accurate and semantically rich image descriptions across both datasets. 

\section{Conclusion and Future Direction}
\label{conclusion}
Combining the Vision Transformer encoder module (ViT) with cross-model attention and the Generative Pre-trained Transformers 4 (GPT 4.0) decoder module can result in an efficient framework, CrossViT-GPT4, for producing visual descriptions. The model uses vision-based feature extraction and language modeling to thoroughly analyze and characterize complex medical images. When textual and visual components are combined, chest X-ray pathology reports can be described and explained more precisely. Transformer-based designs are scalable and adaptable, promoting medical image analysis research and innovation. Their adaptability enables them to efficiently manage diverse datasets and tasks. By enhancing medical image processing, the suggested approach may increase diagnostic precision and assist clinical decision-making platforms. Poor image quality can seriously affect model accuracy by restricting the ability to extract features and generate accurate descriptions. Future research could address this by creating reliable pre-processing methods or incorporating sophisticated denoising and super-resolution techniques to improve performance. Additional research could also increase the applicability and adaptability of the proposed method in a number of scenarios by utilizing a range of large and diverse medical imaging datasets.

\section*{Acknowledgment}
The authors sincerely thank the Cyber Forensics and Malware Analysis Laboratory, Department of IT, DTU, Delhi, India, for providing the necessary resources to complete the research.
\section*{Data Availability Statement:} The paper contains links to all datasets.
\section*{Competing Interests:} The authors declare they have no competing financial interests.
\section*{Conflict of Interest:} The authors declare no conflict of interest.

\bibliography{main}

\begin{thebibliography}{21}
\providecommand{\natexlab}[1]{#1}
\providecommand{\url}[1]{\texttt{#1}}
\providecommand{\urlprefix}{}

\bibitem[{Beddiar et~al.(2023)Beddiar, Djamila-Romaissa and Oussalah, Mourad and Sepp{\"a}nen, Tapio}]{beddiar2023automatic}
Beddiar DR, Oussalah M, Sepp{\"a}nen T.
\newblock Automatic captioning for medical imaging (MIC): a rapid review of literature.
\newblock Artificial intelligence review 2023;56(5):4019--4076.

\bibitem[{Agarwal and Verma(2023)Agarwal, Lakshita and Verma, Bindu}]{agarwal2023comparison}
Agarwal L, Verma B.
\newblock Comparison of Deep Learning Models for Automatic Image Descriptors.
\newblock In: 2023 IEEE 20th India Council International Conference (INDICON) IEEE; 2023. p. 914--919.

\bibitem[{Revathi and Kowshalya(2024)Revathi, BS and Kowshalya, A Meena}]{revathi2024automatic}
Revathi B, Kowshalya AM.
\newblock Automatic image captioning system based on augmentation and ranking mechanism.
\newblock Signal, Image and Video Processing 2024;18(1):265--274.

\bibitem[{Agarwal and Verma(2024)Agarwal, Lakshita and Verma, Bindu}]{agarwal2024methods}
Agarwal L, Verma B.
\newblock From methods to datasets: A survey on Image-Caption Generators.
\newblock Multimedia Tools and Applications 2024;83(9):28077--28123.

\bibitem[{Pan et~al.(2022)Pan, Xuran and Ye, Tianzhu and Han, Dongchen and Song, Shiji and Huang, Gao}]{pan2022contrastive}
Pan X, Ye T, Han D, Song S, Huang G.
\newblock Contrastive language-image pre-training with knowledge graphs.
\newblock Advances in Neural Information Processing Systems 2022;35:22895--22910.

\bibitem[{Sun and Min(2024)Sun, Jing-Tao and Min, Xuan}]{sun2024research}
Sun JT, Min X.
\newblock Research on image caption generation method based on multi-modal pre-training model and text mixup optimization.
\newblock Signal, Image and Video Processing 2024;p. 1--19.

\bibitem[{Yao et~al.(2023)Yao, Ting and Li, Yehao and Pan, Yingwei and Wang, Yu and Zhang, Xiao-Ping and Mei, Tao}]{yao2023dual}
Yao T, Li Y, Pan Y, Wang Y, Zhang XP, Mei T.
\newblock Dual vision transformer.
\newblock IEEE transactions on pattern analysis and machine intelligence 2023;.

\bibitem[{Liu et~al.(2021)Liu, Fenglin and Yin, Changchang and Wu, Xian and Ge, Shen and Zou, Yuexian and Zhang, Ping and Sun, Xu}]{liu2021contrastive}
Liu F, Yin C, Wu X, Ge S, Zou Y, Zhang P, et~al.
\newblock Contrastive attention for automatic chest x-ray report generation.
\newblock arXiv e-print arXiv:210606965 2021;.

\bibitem[{Shaikh and Bharti(2023)Shaikh, Zeeshan and Bharti, Jyoti}]{shaikh2023transformer}
Shaikh Z, Bharti J.
\newblock Transformer-Based Chest X-ray Report Generation Model.
\newblock In: International Conference on Soft Computing and Signal Processing Springer; 2023. p. 227--236.

\bibitem[{Zeiser et~al.(2024)Zeiser, Felipe Andr{\'e} and da Costa, Cristiano Andr{\'e} and de Oliveira Ramos, Gabriel and Maier, Andreas and da Rosa Righi, Rodrigo}]{zeiser2024chexreport}
Zeiser FA, da~Costa CA, de~Oliveira~Ramos G, Maier A, da~Rosa~Righi R.
\newblock CheXReport: A transformer-based architecture to generate chest X-ray reports suggestions.
\newblock Expert Systems with Applications 2024;p. 124644.

\bibitem[{Reale-Nosei et~al.(2024)Reale-Nosei, Gabriel and Amador-Dom{\'\i}nguez, Elvira and Serrano, Emilio}]{reale2024vision}
Reale-Nosei G, Amador-Dom{\'\i}nguez E, Serrano E.
\newblock From vision to text: A comprehensive review of natural image captioning in medical diagnosis and radiology report generation.
\newblock Medical Image Analysis 2024;p. 103264.

\bibitem[{Shahzadi et~al.(2024)Shahzadi, Iqra and Madni, Tahir Mustafa and Janjua, Uzair Iqbal and Batool, Ghanwa and Naz, Bushra and Ali, Muhammad Qasim}]{shahzadi2024csamdt}
Shahzadi I, Madni TM, Janjua UI, Batool G, Naz B, Ali MQ.
\newblock CSAMDT: Conditional Self Attention Memory-Driven Transformers for Radiology Report Generation from Chest X-Ray.
\newblock Journal of Imaging Informatics in Medicine 2024;p. 1--13.

\bibitem[{Chen et~al.(2020)Chen, Zhihong and Song, Yan and Chang, Tsung-Hui and Wan, Xiang}]{chen2020generating}
Chen Z, Song Y, Chang TH, Wan X.
\newblock Generating radiology reports via memory-driven transformer.
\newblock arXiv e-print arXiv:201016056 2020;.

\bibitem[{Wijerathna et~al.(2022)Wijerathna, Vidura and Raveen, Hemaka and Abeygunawardhana, Sachini and Ambegoda, Thanuja D}]{wijerathna2022chest}
Wijerathna V, Raveen H, Abeygunawardhana S, Ambegoda TD.
\newblock Chest x-ray caption generation with chexnet.
\newblock In: 2022 Moratuwa Engineering Research Conference (MERCon) IEEE; 2022. p. 1--6.

\bibitem[{Wang et~al.(2022)Wang, Jun and Bhalerao, Abhir and He, Yulan}]{wang2022cross}
Wang J, Bhalerao A, He Y.
\newblock Cross-modal prototype driven network for radiology report generation.
\newblock In: European Conference on Computer Vision Springer; 2022. p. 563--579.

\bibitem[{Huang et~al.(2023)Huang, Zhongzhen and Zhang, Xiaofan and Zhang, Shaoting}]{huang2023kiut}
Huang Z, Zhang X, Zhang S.
\newblock Kiut: Knowledge-injected u-transformer for radiology report generation.
\newblock In: Proceedings of the IEEE/CVF Conference on Computer Vision and Pattern Recognition; 2023. p. 19809--19818.

\bibitem[{Zeng et~al.(2024)Zeng, Xianhua and Liao, Tianxing and Xu, Liming and Wang, Zhiqiang}]{zeng2024aermnet}
Zeng X, Liao T, Xu L, Wang Z.
\newblock AERMNet: Attention-enhanced relational memory network for medical image report generation.
\newblock Computer Methods and Programs in Biomedicine 2024;244:107979.

\bibitem[{You et~al.(2016)You, Quanzeng and Jin, Hailin and Wang, Zhaowen and Fang, Chen and Luo, Jiebo}]{you2016image}
You Q, Jin H, Wang Z, Fang C, Luo J.
\newblock Image captioning with semantic attention.
\newblock In: Proceedings of the IEEE conference on computer vision and pattern recognition; 2016. p. 4651--4659.

\bibitem[{Jing et~al.(2017)Jing, Baoyu and Xie, Pengtao and Xing, Eric}]{jing2017automatic}
Jing B, Xie P, Xing E.
\newblock On the automatic generation of medical imaging reports.
\newblock arXiv e-print arXiv:171108195 2017;.

\bibitem[{Yan and Pei(2022)Yan, Bin and Pei, Mingtao}]{yan2022clinical}
Yan B, Pei M.
\newblock Clinical-bert: Vision-language pre-training for radiograph diagnosis and reports generation.
\newblock In: Proceedings of the AAAI Conference on Artificial Intelligence, vol.~36; 2022. p. 2982--2990.

\bibitem[{Ouis and Akhloufi(2024)Ouis, Mohammed Yasser and Akhloufi, Moulay A}]{ouis2024chestbiox}
Ouis MY, Akhloufi MA.
\newblock ChestBioX-Gen: contextual biomedical report generation from chest X-ray images using BioGPT and co-attention mechanism.
\newblock Frontiers in Imaging 2024;3:1373420.

\end{thebibliography}
\end{document}